\documentclass[sigconf]{acmart}
\usepackage{xcolor}

\usepackage[english]{babel}
\usepackage{xcolor}

\usepackage{amsmath}
\usepackage{comment}
\usepackage{graphicx}
\newcommand{\YZ}[1]{{\color{blue} \sf (YZ: #1)}}

\usepackage{xcolor}
\usepackage[most]{tcolorbox}

\title{Diagnosing and Resolving Android Applications Building Issues: An Empirical Study}
\author{Lakshmi Priya Bodepudi}
\affiliation{
  \institution{University of Cincinnati}
  \city{Cincinnati}
  \state{Ohio}
  \country{USA}
}
\email{bodepula@mail.uc.edu}

\author{Yutong Zhao}
\affiliation{
  \institution{California State University, Long Beach}
  \city{Long Beach}
  \state{California}
  \country{USA}
}
\email{Yutong.Zhao@csulb.edu}

\author{Ming Quan Fu}
\affiliation{
  \institution{University of Central Missouri}
  \city{Warrensburg}
  \state{Missouri}
  \country{USA}
}
\email{mingfu@ucmo.edu}

\author{Yuanyuan Wu}
\affiliation{
  \institution{University of Cincinnati}
  \city{Cincinnati}
  \state{Ohio}
  \country{USA}
}
\email{wu3yy@mail.uc.edu}

\author{Sen He}
\affiliation{
  \institution{The University of Arizona}
  \city{Tucson}
  \state{Arizona}
  \country{USA}
}
\email{senhe@arizona.edu}

\author{Yu Zhao}
\affiliation{
  \institution{University of Cincinnati}
  \city{Cincinnati}
  \state{Ohio}
  \country{USA}
}
\email{zhao3y3@ucmail.uc.edu}

\begin{document}
\begin{abstract}

Building Android applications reliably remains a persistent challenge due to complex dependencies, diverse configurations, and the rapid evolution of the Android ecosystem. This study conducts an empirical analysis of 200 open-source Android projects written in Java and Kotlin to diagnose and resolve build failures. Through a five-phase process encompassing data collection, build execution, failure classification, repair strategy design, and LLM-assisted evaluation, we identified four primary types of build errors: environment issues, dependency and Gradle task errors, configuration problems, and syntax/API incompatibilities. Among the 135 projects that initially failed to build, our diagnostic and repair strategy enabled developers to resolve 102 cases (75.56\%), significantly reducing troubleshooting effort. We further examined the potential of Large Language Models, such as GPT-5, to assist in error diagnosis, achieving a 53.3\% success rate in suggesting viable fixes. An analysis of project attributes revealed that build success is influenced by programming language, project age, and app size. These findings provide practical insights into improving Android build reliability and advancing AI-assisted software maintenance.


\end{abstract}

\maketitle

\section{Introduction}

The popularity of mobile applications has grown rapidly in recent years. Currently, the Google Play Store hosts approximately three million applications~\cite{googleplaylink}, making mobile app development an integral part of modern software engineering. Meanwhile, millions of Android application source codes are hosted on public code repositories such as GitHub~\cite{Github} and Bitbucket~\cite{bitbucket}, primarily written in Java and Kotlin. These applications are typically developed using the Android Software Development Kit (SDK) and build tools such as Gradle~\cite{liu2024understanding}. Developers commonly use Android Studio~\cite{craig2015learn} to automate the process from source code to deployable application. However, build failures remain a widespread challenge in Android development, significantly hindering efficient development and continuous delivery~\cite{madeja2021automating, liu2024understanding, zhang2016android}.

Recent studies have shown that only about 31.32\% of open-source Android applications hosted on GitHub can be successfully built automatically~\cite{liu2024understanding}. This low build success rate not only reduces development efficiency but also limits large-scale testing, analysis, and utilization of Android applications. Therefore, improving build success rates is essential for supporting sustainable growth of the Android ecosystem and enabling large-scale software engineering research and practice.

Prior studies~\cite{sulir2016quantitative, sulir2020large, hassan2017automatic} have examined build\ failures in Java projects, but their applicability to modern Android builds remains limited. In contrast, some studies have specifically focused on Android projects. Jha et al.~\cite{jha2017developer} conducted a large-scale empirical analysis of errors in Manifest files, but their work was limited to configuration-level issues and did not address more complex factors such as multiple programming languages, third-party libraries, or diverse dependency sources. Liu et al.~\cite{liu2024understanding} analyzed the evolution and quality of Android build systems at scale and identified five major root causes of build failures. However, because their study relied mainly on automated build experiments, it may have overlooked semantic-level issues such as API or SDK compatibility problems.

Beyond identifying root causes, several key areas remain underexplored: (1) the design of a systematic diagnostic and repair strategy, (2) the use of large language models (LLMs) to support build processes, and (3) an in-depth understanding of how programming language, project characteristics, and development period contribute to project attributes associated with build failures.

In this study, we conducted an empirical analysis of 200 Android applications collected from GitHub between 2012 and 2025, investing approximately 1,000 hours in manual diagnosis to identify root causes of build failures and develop effective repair strategies. The results show that 135 of these projects failed to build directly in Android Studio~\cite{craig2015learn}. We categorized the root causes of these failures into four major types: environment issues (45\%), dependency and Gradle task errors (42\%), configuration errors (8\%), and syntax/API errors (5\%). The primary causes include outdated libraries, misconfigured Gradle files, Java–Kotlin interoperability issues, frequent updates to the Android Gradle Plugin, and the lack of standardized development environments.

Based on these findings, we designed a diagnostic and repair strategy practical guidelines to address specific build errors: (1) Incompatibility Issues Caused by Gradle, (2) Java version conflicts, (3) Deprecated Gradle plugins, (4) Corrupted Gradle cache or daemon, (5) Manifest \& AndroidX issues, (6) Missing or broken dependencies, (7) Code-level cleanup, and (8) SDK/API level mismatch. These guidelines aim to help developers efficiently identify and resolve build issues. Projects were classified into four categories according to the level of effort required for repair: no issues, minor failures, major failures, and unresolvable failures. Ultimately, applying these guidelines, we successfully repaired 92 minor issues projects (46\%) and 10 major issues projects (5\%), fixing a total of 102 build issues.

We also leveraged LLMs such as GPT-5~\cite{GPT4o}, inspired by recent progress in software engineering applications~\cite{wang2023empirical, ju2024study, wen2024autodroid}, to assist in diagnosing and repairing build failures. Our findings indicate that LLM recommendations facilitated the resolution of 53.3\% of identified errors. However, for 13\% of errors that were fixable by human intervention, GPT-5 failed to generate correct and actionable repair responses. Moreover, for projects that could not be built by any developer, LLMs also failed to provide meaningful or improved solutions. These results demonstrate significant potential for AI-assisted workflows to enhance Android build efficiency and reliability, yet also reveal current limitations that warrant further improvement.

Further analysis of 200 Android projects revealed that project attributes significantly affect build success. Programming language played an important role: Kotlin projects exhibited higher build stability, while Java projects were more prone to environment and dependency-related failures. Project age is another project attribute, with older apps relying on deprecated APIs or obsolete build tools being harder to compile and fix. App size also influenced outcomes, as larger projects with more dependencies tended to experience more build failures. These findings highlight how language choice, project characteristics, and development history jointly impact Android build reliability, providing guidance for prioritizing apps in testing and maintenance workflows.

All repair processes were documented with detailed logs, screenshots, and step-by-step procedures, and the dataset has been made publicly available~\cite{thisproject}. To the best of our knowledge, this is the first empirical study to release an Android build trace dataset that includes both successful and failed builds, providing a reproducible foundation for future research on the diagnosis and repair of Android build failures.



This study makes key contributions toward advancing the understanding and improvement of Android applications' buildability.
\begin{enumerate}
\item This work performs a detailed analysis of Android build failures and proposes the first systematic strategy to diagnose and resolve them efficiently.
\item It evaluates the effectiveness of LLMs in error diagnosis and solution generation for Android build failures, serving as an aid to human developers.
\item It presents the first empirical study examining how Android build failures relate to programming language, project size, popularity, and temporal evolution.
\item It constructs and releases the first open dataset of 200 Android apps with detailed logs, screenshots, and solutions, enabling future reuse and reproducibility~\cite{thisproject}.
\end{enumerate}



\section{Research Questions}

\noindent
\textbf{RQ1: What are the categories of issues encountered during compilation of open-source Android applications?}\\
This question aims to systematically classify build failures in open-source Android applications.
We built 200 GitHub apps using \textit{Android Studio Ladybug Feature Drop 2024.2.2} to develop a concise taxonomy, quantify failure types, and identify root causes.
We hypothesize that most failures fall into four categories --- environment, dependency and Gradle, configuration, and syntax/API --- and that distinct patterns can be observed across these categories.

\noindent
\textbf{RQ2: How can practitioners design an efficient and repeatable strategy to resolve these compilation issues?}
This question focuses on creating a structured, repeatable workflow to resolve build failures with minimal trial-and-error.
By analyzing error logs and fixes from 200 Android apps, we define a multi-step process along with eight specific guidelines for error detection and resolution.
We hypothesize that standardizing these steps will significantly enhance repair efficiency and streamline build methodologies~ across projects of different sizes and complexities. 

\noindent
\textbf{RQ3: To what extent can LLMs assist in the diagnosis and resolution of Android build failures?} To evaluate the effectiveness of LLMs in Android build failure resolution, we conducted a controlled experiment with \textit{GPT-5}.
We selected 15 representative build failures, including five minor, five major, and five unfixable cases, categorized according to their fixing difficulty of repair by human developers. For each case, we provided the raw error logs along with designed prompts and asked the model to suggest potential fixes. The model’s effectiveness was assessed based on the success rate, defined as the proportion of cases in which the recommended solution resulted in a successful build.

\noindent
\textbf{RQ4: What project attributes influence the build of Android applications?} This question investigates the key project attributes correlated with Android build success by examining project attributes (e.g., programming language, project size, rating, years). Specifically, we analyze how language choice (Java vs. Kotlin) affects error types and how project size correlates with build failure rates and repair success, contrasting large applications with smaller ones. The goal is to provide a comprehensive project attributes analysis that clarifies how these attributes collectively influence build success across a diverse set of modern Android projects.

\section{Study Subjects and Approach}

\subsection{Study Subjects}
\label{sec:subjects}

\noindent
To investigate the evolution of Android build reliability, we constructed a dataset of 200 open-source applications from GitHub. The projects, created between 2012 and 2025, include applications written in Java, Kotlin, or a combination of both. We employed GitHub’s advanced search filters to select a representative and heterogeneous sample. The filters are based on criteria, including the programming language, creation date, and popularity. The intentional diversity in project age and language is crucial for capturing the wide spectrum of build failures that have emerged alongside the evolution of Android's tooling, libraries, and development conventions.

\noindent
To ensure the diversity and relevance of our dataset, we followed the following five criteria:
\begin{itemize}
    \item \textbf{Language:}
    {Applications must be written in \textit{Java}, \textit{Kotlin}, or a combination of both, the official languages for Android development.}

    \item \textbf{Domain:}
    {Applications are drawn from various categories, such as productivity, media, security, and utilities, to ensure broad domain representation.}
    
    \item \textbf{Accessibility:}
    {All applications are open-source and hosted on GitHub, ensuring our study is fully reproducible.}

    \item \textbf{Time spanning}:
    {Applications feature development activity between 2012 and 2025, capturing a wide range of toolchain compatibility challenges.}
    
    \item \textbf{Build system:}
    {Applications must use the Gradle build system to allow for a standardized compilation process.}
\end{itemize}


\noindent 
For our experimental setup, each project was imported into \textit{Android Studio Ladybug Feature drop | 2024.2.2}. We configured a standardized build environment using the latest Gradle plugin (v8.8.0), modern Java and Kotlin SDKs, and dependencies from Maven Central and Google. This consistent setup was essential for systematically identifying build failure patterns and analyzing the evolving challenges in Android open-source development.

\subsection{Study Approach}
\label{sec:approach}

\begin{figure}[ht]
\centering
 \includegraphics[scale=0.7]{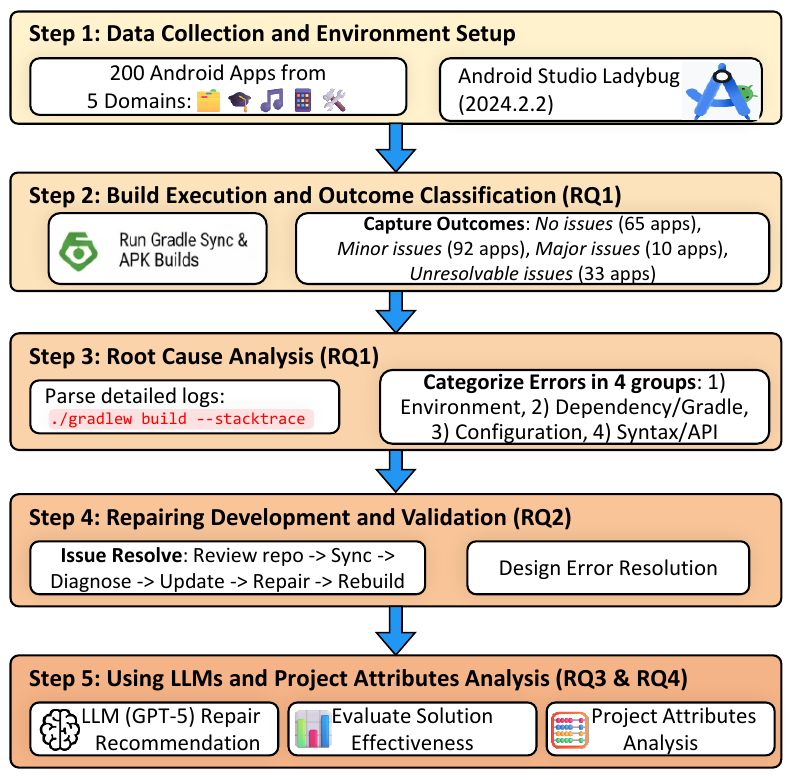} 
  \caption{Study Approach Overview} 
  \label{fig:overview}
\end{figure}

\noindent
Our study approach consists of five sequential steps, as depicted in the workflow overview shown in Fig.~\ref{fig:overview}.

\begin{itemize}
  \item \textbf{Step 1: Data Collection and Environment Setup.}
        We collected 200 open-source Android applications from GitHub, spanning five domains: file management (e.g., \textit{Amaze File Manager}), education (e.g., \textit{BookFinder}), media streaming (e.g., \textit{MultimediaApp}), messaging (e.g., \textit{SMS Backup+}), and utilities (e.g., \textit{Aegis Authenticator}). For each repository, we cataloged metadata, including name, language, year, URL before cloning and importing it into \textit{Android Studio Ladybug (2024.2.2)}. The build environment was configured to support multiple Java (8, 11, 17) and Kotlin (1.8.20, 1.9.0) versions, targeting Android SDKs from API 21 to 34.



\item \textbf{Step 2: Build Execution and Outcome Classification (RQ1).}  We ran Gradle synchronization and APK builds for each application, capturing detailed build logs, a process that took 10 to 15 minutes per build. Initial outcomes were classified based on the raw error logs from \textit{Android Studio} and later refined through the diagnostic analysis described in Step 3 and 4. Outcomes were categorized as follows: 
    \begin{enumerate}
      \item \emph{No issues} (65 apps); 
      \item \emph{Minor issues} resolvable with simple configuration tweaks (92 apps);
      \item \emph{Major issues} requiring substantial debugging or refactoring (10 apps);
      \item \emph{Unresolvable issues} (the failures which cannot be fixed) due to deprecated tools or missing resources (33 apps). 
    \end{enumerate}
   
\item \textbf{Step 3: Root-Cause Analysis (RQ1)}
    We analyzed all logged errors and grouped them into four primary categories with their root causes: environment misconfigurations, dependency and Gradle task issues, configuration errors, syntax/API mismatches. Based on this taxonomy, we examined all error logs from the unsuccessful builds in detail. Each log was inspected line by line to trace errors back to their origins—whether they stemmed from environment misconfigurations, outdated components, or code-level issues.


  \item \textbf{Step 4: Repairing Development and Validation (RQ-2).} \\
Guided by our taxonomy, we created and tested a six-step workflow for resolving issues: (1) reviewing the repository, (2) syncing the project, (3) running diagnostic tests, (4) updating standard versions, (5) making targeted repairs, (6) and rebuilding. We also summarized eight distinct error types and solutions, categorized under Step 2's four solution categories. For each error, we provided detailed detection methods, repair solutions, and examples for resolution. We used this process on 102 common build failures to check its effectiveness. For each case, we tracked the time needed, whether the fix worked, and how difficult it was. We classified failures as \textit{minor issues} if only small changes are required such as java or gradle versions need to be upgraded and \textit{major issues} if it requires various changes such as  (e.g., library migrations, significant refactoring). All steps and results were documented to check the workflow's validity.

\begin{table*}[!th]
\caption{Summary of Root Causes for Compilation Failures}
\label{tab:android_errors}
\centering
\scalebox{0.85}{
\begin{tabular}{|p{3cm}|c|c|p{13cm}|}
\hline

\textbf{Error Type} & \textbf{Occurrence} & \textbf{Percentage} & \textbf{Description} \\
\hline
Development Environment Errors & 62& 45\%& Local setup mismatches such as unsupported JDK/SDK versions, missing Android SDK components, or invalid environment variables.\\
\hline
Dependency and Gradle task errors      & 58& 42\%& Failures in library resolution due to version mismatches, missing repositories, or conflicting transitive dependencies. Build failures caused by missing or misconfigured Gradle tasks or plugins, including syntax errors in custom tasks.\\
\hline
Configuration errors  & 12& 8\%& Incorrect settings in \texttt{build.gradle}, \texttt{AndroidManifest.xml}, or \texttt{gradle.properties} (e.g., misconfigured \texttt{minSdkVersion}, missing \texttt{android:exported} attribute).\\
\hline
Syntax/API errors & 7& 5\% & Compilation errors from deprecated or missing Android/third-party APIs, or incorrect or missing Kotlin/Java syntax.\\
\hline
\textbf{Total} & 139& 100\% & Summary of all categorized errors.\\
\hline
\end{tabular}
}
\end{table*}

\item \textbf{Step 5: Using LLM'S and Project Attribute Analysis (RQ-3 \& RQ-4).} 
We did a controlled experiment to see how well LLMs GPT-5 work for creating failure resolution. A cutting-edge LLM was given raw error logs from a stratified sample of build failures and told to come up with useful repair plans. The main measure was how well the LLM-recommended solutions worked. At the same time, we did a longitudinal analysis on the whole dataset to find links between program features (such as language, development year, ratings, and project size).
\end{itemize}
 
By executing these five steps, we establish a comprehensive and reproducible methodology to identify, classify, and resolve build failures in a diverse set of Android applications. The resulting build failure resolution strategy and project attribute analyzes offer both practitioners and researchers practical guidance on common compilation challenges and effective resolution strategies.



\section{Study Results}
\label{sec:results}

\subsection{\textbf{RQ1: {Categories of issues in Android applications building}}}
\label{sec:rq1}

\noindent 

For this research question, we aimed to analyze the root causes of build failures to summarize the categories of issues in Android applications. We built all 200 selected apps in our dataset and examined the error logs to identify their root causes. In the initial build using Android Studio, only 65 applications (32.5\%) compiled successfully without modification, while the remaining 135 (67.5\%) failed, highlighting the prevalence of build issues in open-source Android projects. We focused on these 135 failed applications to analyze their logs and categorize the types of issues.

To ensure the accuracy of our root cause analysis, we manually resolved these issues and used the success of these resolutions to verify whether the identified root causes were correct. Among these failures, 102 applications (51\% of the total dataset, or 75.6\% of all failures) were successfully resolved through targeted code or configuration changes. The remaining 33 applications (16.5\% of the dataset, or 24.4\% of all failures) were deemed unfixable due to abandoned codebases, missing source files, or irreparable dependency conflicts. Our subsequent root-cause analysis focuses on the 102 apps that initially failed to build but could be successfully resolved.

Based on the 102 apps from the resolved issues, we analyzed the aggregated error logs and categorized compilation failures into four root-cause groups as described in Table~\ref{tab:android_errors}: 1) Development Environment Configuration Errors (45\%), 2) Dependency Management Errors (42\%), 3) Configuration Errors (8\%), and 4) Syntax/API Errors (5\%). These percentages are based on all failures.

Note that while issues are grouped into four categories, one app may suffer muliple issues that span multiple groups. Independent simultaneous issues can therefore be counted in multiple categories. Thus, the total issue count (139) exceeds the 102 problematic apps fixed out of 200 tested (see Table~\ref{tab:android_errors}).

\noindent
{\textbf{1. Development Environment Issues (Environment Errors)}}

\noindent Environment configuration errors accounted for 45\% of all build failures, primarily caused by incompatibilities between the local environment and external tooling (e.g., JDK/Gradle, environment variables) rather than code defects.
\textit{TajMahalAppAndroid} failed to build with JDK 8 and Gradle 8.9, but upgrading to JDK 17 resolved the issue within 20 minutes; \textit{PetCareApp} completed its migration to Java 17 in 20 minutes.
Common error messages included ``\texttt{Unsupported Java version}'' and ``\texttt{Android SDK path not found}'', with legacy Java projects (e.g., \textit{Drawingapp}) more prone to such issues.
Typical fixes involved aligning JDK/SDK versions, correcting environment variables (e.g., \verb|ANDROID_HOME|), clearing the Gradle cache, and re-syncing the project, usually within 20 to 30 minutes.
Standardizing and regularly validating development environment configurations is critical for preventing these avoidable build interruptions.

\noindent
\textbf{2. Dependency and Gradle Task Errors}\\
\noindent
Dependency errors accounted for 42\% of all build failures, mainly caused by version mismatches, missing repositories, or transitive dependency conflicts.
For example, in the \textit{Meme Share} app, \textit{Volley 1.1.1} triggered a “\texttt{cannot resolve symbol}” error, which was fixed by upgrading to \textit{1.2.1} and adding the \texttt{mavenCentral()} repository.
Age Calculator required upgrading the Gradle wrapper, while openNoteScanner resolved issues by replacing the deprecated dependency “us.feras.mdv:markdownview:1.1.0." with\\ \texttt{ com.github.falnatsheh:MarkdownView:58e6298f38
}. 

These fixes typically took 30–40 minutes, involving log inspection and repository checks. Java projects (e.g., \textit{SMS Backup+}) showed dependency errors, often related to legacy libraries. The main resolution steps included updating \texttt{build.gradle}, re-syncing Gradle, and ensuring dependency versions aligned with the project’s API level. In some cases, unstable network conditions can further corrupt Gradle cache and daemon processes, leading to synchronization failures during dependency resolution.

Another common failure type was Gradle task errors, which stemmed from missing or misconfigured tasks, plugins, or syntax issues rather than dependency errors such as library conflicts. For instance, the build failure of \textit{InstaMaterial} was because of an \texttt{Gradle and AGP} error, which was resolved by correcting the task; Omni Notes resolved execution errors by upgrading AGP. Fixing similar issues in complex projects (e.g., \textit{LikeAnimation} ) typically took 15–20 minutes. Kotlin projects (e.g., \textit{LibreTube}) experienced no errors, likely due to stricter validation.

\noindent
\textbf{3. Configuration Errors:} \\
\noindent
Configuration errors accounted for about 8\% of all build failures, mainly caused by incorrect settings in \texttt{build.gradle}, \\ \texttt{AndroidManifest.xml}, or \texttt{gradle.properties}. Common issues included build type misconfigurations, missing permissions, and invalid SDK declarations. For example, \textit{Cryptography} failed on Android 12+ (API 31) due to a missing android:exported attribute, which was fixed in about 20 minutes; \textit{Basic Room Database Master} required updating minSdkVersion; and \textit{CalculatorApp} needed a clean rebuild due to corrupted cache from misconfiguration. Java projects (e.g., \textit{Wallpaper}) showed more configuration errors than Kotlin ones, often linked to legacy settings. While Insta Material fixed a missing google() repository in 20–30 minutes. Common error logs included ``\texttt{Manifest merger failed}'' and ``\texttt{Invalid minSdkVersion}''. These findings highlight the importance of regularly checking configuration files to prevent build failures and reduce debugging time.\\
Dependency errors come from external libraries, while configuration errors come from internal project settings, separating them makes troubleshooting faster and more accurate.

\noindent
\textbf{4. Syntax/API Errors:} \\
\noindent
 Syntax and API errors accounted for 5\% of all build failures, primarily resulting from the use of deprecated Android or third-party APIs, as well as incorrect Kotlin or Java syntax. For example, in the \emph{Bible App}, deprecated \textit{SQLite} calls were resolved by checking the database libraries and changed the versions. Likewise, the \emph{Voice Record App} project required omitting the unwanted syntax so that it gets compiled successfully. Error logs for these issues commonly displayed messages such as ``\texttt{NoSuchMethodError}'', which clearly indicates that a method call in the code is targeting an outdated or removed API. These issues were more common in Kotlin projects (e.g., \emph{Book Finder}) due to stricter type-checking and language-level safety constraints.

\begin{tcolorbox}[size=title, opacityfill=0.15, enhanced, breakable]
\textbf{RQ1 Summary}: Among the 200 open-source Android applications, 32.5\% built successfully without intervention, 46\% were fixed with minor effort (20–30 min), 5\% required major refactoring (4–8 h), and 16.5\% were unfixable due to abandoned or corrupted code.
Environment misconfigurations (45\%) and dependency management errors (42\%) were the primary causes, followed by configuration errors (8\%) and syntax/API mismatches (5\%). Java projects showed higher dependency and environment error rates than Kotlin projects.
These results suggest that automated dependency and manifest validation can prevent most build failures, while the unfixable and high-effort cases highlight the need for continuous code modernization and improved tooling.
\end{tcolorbox}
\subsection{\textbf{RQ2: Diagnostic methods and practical resolution guidelines}}
\label{sec:rq2}

\noindent

In this research question, we focus on designing guidelines and diagnosing and resolving build issues. Within the four primary error categories identified in RQ1, we observed that some categories may require multiple diagnostic and resolution strategies due to recurring patterns of specific failures. To provide targeted and practical guidance for practitioners, this section details eight frequently recurring sub-issues across the four primary error categories. It is important to note that these sub-issues are not new categories; rather, they are concrete manifestations of the broader root causes discussed previously, such as Dependency and Configuration Errors. For each sub-issue, we present a focused detection and resolution strategy designed to improve build reliability and developer productivity.


\vspace{10 pt}
 
\noindent
\textbf{1. Incompatibility Issues Caused by Outdated Gradle and Android Gradle Plugin (AGP) Versions (Gradle Task Error)}
\hspace{5pt}

{\textbf{ISSUE}}- This issue arises when the version of Gradle used in a project is incompatible with the version of the Android Gradle Plugin (AGP) it depends on.
Each AGP version is designed to work with a specific range of Gradle versions. If they don’t align, the build process may fail or produce various unexpected errors.
It’s much like trying to open a modern file format with outdated software — the old program simply can’t recognize or handle it.
\hspace{5pt}

{\textbf{DETECTION}} -
 Detection is typically straightforward, as error logs often explicitly state the incompatibility. 
A typical error message (with version numbers varying by case) is as follows:
The project is using an incompatible version of the Android Gradle Plugin (AGP 8.11.1). According to the latest feature documentation of Android Studio Ladybug 2024.2.2, the highest supported version is 8.8.0.
\hspace{-5pt}

{\textbf{SOLUTION}}- To solve this problem, we updated Gradle to the highest compatible version indicated in the error log (this version may vary depending on the project).
For the AGP compatibility issue, we set the version to 8.8.0, as this version is fully compatible with Android Studio Ladybug 2024.2.2.
The main changes were made in the build.gradle file, and if necessary, also in the gradle-wrapper.properties file.
The entire fix took about 20–30 minutes, and the issue was confirmed resolved after a successful project synchronization.
\hspace{-5pt}


{\textbf{EXAMPLE}} - While compiling the \emph{Android-Studio-Tutorials-Java-Edition}\cite{Android} application, we encountered this AGP incompatibility. The project originally used version 8.11.1, which is not compatible with the current version of Android Studio (\textit{Ladybug 2024.2.2}). We resolved this by downgrading the AGP version to 8.8.0 in the configuration file, which allowed the project to compile successfully.

\vspace{10 pt}
 
\noindent
\textbf{2. Java Version Conflicts (Environment Issue)}

{\textbf{ISSUE}}-  Android projects typically need to be built with a specific Java version.
If the installed Java version (such as 8, 11, or 17) is different from the one required by the project, conflicts may occur.

Such conflicts can lead to compilation errors or runtime incompatibilities because certain Java features may be missing or behave differently\ across versions.

{\textbf{DETECTION}} -  Error logs typically report a JVM version incompatibility or Gradle JVM mismatch. A common error message pattern is \texttt{"Unsupported Java. Your build is currently configured to use Java 1.8 and Gradle 8.x, which requires Java 17 or newer."}

\textbf{SOLUTION} -  The resolution involved reconfiguring the build environment. Specifically, we navigated to the project's Gradle settings within Android Studio and selected the correct JVM version required by the project (e.g., Java 11 or 17). After applying this change, a project re-sync resolved the issue. This fix typically took approximately 15-20 minutes. 

{\textbf{EXAMPLE}} - While compiling the Age for Minutes kt application, we encountered an error indicating that the current AGP requires a Gradle JVM that supports Java 11.
To resolve this issue, we upgraded Java to version 11, and the project was compiled successfully afterward.\cite{Age}

\vspace{10 pt}

\noindent
\textbf{3. Outdated/Deprecated Gradle Plugins (Dependency Error)} 

{\textbf{ISSUE}}- Gradle heavily relies on external plugins to perform tasks such as building APKs and managing dependencies.
Over time, some of these plugins may be replaced or become unmaintained.
If a project continues to reference outdated plugins, Gradle will be unable to execute the related tasks properly, which may lead to build failures, missing functionality, or runtime errors or changing the keywords.

{\textbf{DETECTION}}-  An example of the error message is as follows (the version number may vary depending on the project):
``The specified Gradle installation directory (/Applications/Android \\ Studio.app/Contents/gradle/gradle-2.10)  does not exist.”, “compile is deprecated. Please replace it with implementation.'' The actual version number may differ across projects, but the message and the underlying issue remain essentially the same.

{\textbf{SOLUTION}}- The solution includes updating the gradle version to the compatible android gradle plugin and replacing some keywords which are not used in the new gradle versions such as compile, testCompile, and androidTestCompile are changed to implementation, testImplementation, and androidTestImplementation within the build.gradle file.
This lasted between 20-30 minutes.By reviewing the build logs for warnings and errors related to deprecated plugins or outdated keywords, the issue can be quickly identified and resolved. 

{\textbf{EXAMPLE}} - In \textit{Like Animation} app~\cite{Anime}, after upgrading the Gradle and AGP versions, we encountered an error caused by deprecated \texttt{compile} keywords, which was resolved by replacing them with \texttt{implementation}. The project then compiled successfully.

\vspace{10 pt}

\noindent
\textbf{4. Corrupted Gradle Cache or Daemon (Network Download Latency Error)} 

{\textbf{ISSUE}}- Gradle stores build files and metadata in a local cache to accelerate subsequent builds. However, this cache or Gradle's background process (the Daemon) may become corrupted --- often due to unstable network connections, interrupted builds, system updates, or manual file deletions --- leading to unexpected build failures. In most cases, performing a cache clean-up or restarting the Daemon resolves the issue.


{\textbf{DETECTION}}- It shows messages in the log such as ``\texttt{Gradle's dependency cache may be corrupt}'' (this sometimes occurs after a network connection timeout.)

\textbf{SOLUTION}- The solution involved clearing the Gradle cache by running \texttt{./gradlew clean}, restarting the daemon processes, and rebuilding the project with \texttt{./gradlew clean build}. An alternative solution employed was using the ``Invalidate Caches / Restart'' option in Android Studio, which cleared cached data and forced dependency re-downloads. This approach refreshed the environment and resolved the issue within 15–30 minutes. This step is typically necessary when compiling projects with updated Gradle or dependency versions to ensure the outdated cache is fully removed.

{\textbf{EXAMPLE}}-  While compiling the \textit{CalculatorApp}~\cite{Calculator}, multiple errors occurred during the Gradle update and sync process. One of the possible causes indicated was a corrupted Gradle cache. To resolve the issue, we cleared all caches and restarted the project, after which the application compiled successfully. 

\vspace{10 pt}

\noindent
\textbf{5. Manifest \& AndroidX Issues (Configuration Error)} 

{\textbf{ISSUE}}- The Android Manifest defines key app details (like permissions and activity declarations). Errors happen when this file doesn’t align with project dependencies or when the app mixes old Android support libraries with newer AndroidX packages. These mismatches lead to runtime or build-level conflicts, often surfacing as ``\texttt{class not found}'' or namespace errors.


{\textbf{SOLUTION}}- The resolution for this section involved two distinct actions. For the Manifest issue, we first addressed the new Android 12 (API 31) requirement: any app targeting SDK 31+ must explicitly declare \texttt{android:exported} for components with \texttt{<intent-filter>} elements. We resolved this by adding \texttt{android:exported="true"} to the relevant \texttt{<activity>} tag in the \texttt{AndroidManifest.xml} file and resyncing the project. For AndroidX conflicts, the solution was to enable AndroidX in the \texttt{gradle.properties} file by setting \texttt{android.useAndroidX=true} and ensuring all legacy support libraries were migrated. This fix took approximately 20 minutes.

 {\textbf{EXAMPLE}}- In the \emph{Cryptography}~\cite{Cryptography} app, the build failed with an error indicating a missing attribute in the manifest file. We added \texttt{android:exported="true"} to the main activity, after which it successfully compiled.

\vspace{10 pt}

\noindent
\textbf{6. Missing or Broken Dependencies (Dependency Error)} 

{\textbf{ISSUE}}- Every Android project relies on external libraries. If one of these dependencies is missing, removed from its online repository, or declared incorrectly, the build system can’t find it. Think of it like referencing a book in a library that’s no longer on the shelf—the process halts because the resource doesn’t exist. 

{\textbf{DETECTION}}- This error occurs when Gradle fails to resolve project dependencies.
Common causes: wrong coordinates or versions, network/repo issues, or removed/unmaintained third-party libraries.
Typical logs include ``\texttt{Could not determine dependencies}'' or ``\texttt{Could not resolve all files}''.

{\textbf{SOLUTION}}- For missing or broken dependencies, we identified and updated them to the latest stable versions in the \texttt{build.gradle} file, re-synced the project, and the build completed successfully.


{\textbf{EXAMPLE}}- In the \textit{Meme Share App}~\cite{Meme}, we found that the Volley dependency was outdated. We updated its version in the \texttt{build.gradle} file from 1.1.1 to 1.2.1, after which the project compiled successfully. In the \textit{Water Reminder App}~\cite{water}, upgrading the Java version initially introduced several dependency errors. We resolved these by updating the affected libraries to their latest stable versions and commenting out the broken dependencies that continued to cause build failures.


\vspace{10 pt}

\noindent
\textbf{7.Code-Level Cleanup and Recompilation (Syntax/API Error)} 

{\textbf{ISSUE}}- This error occurs when, after updating the application's dependencies, certain code segments become incompatible because the referenced libraries have been modified or removed. Build logs may display messages about missing resources, deprecated functions, or incompatible APIs, particularly in core components such as the main activity. These errors indicate that the source code references elements that no longer exist or whose behavior has changed in the updated library versions.



{\textbf{DETECTION}}- The build log fails with compilation errors, such as ``\texttt{Android resource linking failed}''. More commonly, the IDE itself flags errors directly in Java/Kotlin source code, such as ``\texttt{Cannot resolve symbol}'' for a removed class or type-mismatch errors for a changed method signature.

{\textbf{SOLUTION}}-  The correct solution is not to simply remove or comment out the broken code, as this would remove functionality. Instead, the code must be refactored to adapt to the new library's API. This involves identifying the new, correct method calls or classes to use in place of the deprecated or removed ones. After refactoring the source code, we recompiled the project. This process typically took 30 to 45 minutes.

{\textbf{EXAMPLE}}- In \textit{Voice Record App}\cite{Voice} we added one new dependency and the unwanted syntax in the activity main xml file is omitted so that the app gets compiled successfully and this happens only if you remove the unwanted code(cleanup) related to the broken dependencies. 

\vspace{10 pt}

\noindent
\textbf{8. SDK/API Level Mismatch or Upgrade Incompatibility   (Configuration Error)} 

{\textbf{ISSUE}}- Each Android API level represents a different version of the operating system. If an app targets an outdated SDK or references APIs not supported in the configured target version, the build can fail. This usually occurs when the app’s build.gradle file doesn’t align with the installed SDK versions, leading to errors like ``\texttt{resource not found}'' or ``\texttt{method undefined}''. 

{\textbf{DETECTION}}– This type of error usually occurs when the project’s compileSdkVersion is too low and does not meet the Android API level required by the dependency libraries.
Common scenarios include:
The dependency library (either third-party or official) requires a higher Android API level (e.g., 34 or above);
The project’s current compileSdkVersion is lower than the minimum required version;
The Android Gradle Plugin version being used may also limit the maximum recommended compileSdkVersion.
When such a mismatch occurs, Gradle will report an error in the build logs, indicating that the current compileSdkVersion does not meet the requirements of the dependency.

{\textbf{SOLUTION}}– Update the project to use compileSdkVerion of at least 34 in build.gradle . This takes 20 minutes for fixing and aligning with the specified SDK version (e.g., API 33). 

{\textbf{EXAMPLE}}- While building this app Using Basic Room Database Master\cite{using} after updating the java version it gave error for the sdk version and then we updated it to version 34 from 30 and then it worked.  


\begin{tcolorbox}[size=title, opacityfill=0.15, enhanced, breakable]
\textbf{RQ2 Validation}:
Applying our build failure resolution strategy to 200 apps revealed that 32.5\% were built successfully without intervention, while 67.5\% failed initially.
We repaired 46\% with minor fixes (1–2h) and 5\% with major refactoring (4–6h), leaving 16.5\% unfixable due to fragmented or abandoned codebases.
These results confirm the effectiveness of our strategy in addressing both simple and complex build issues.

\end{tcolorbox}



\subsection{\textbf{RQ3: To what extent can Large Language Models assist in the diagnosis and resolution of Android build failures?}}

This research investigates whether Large Language Models (LLMs) can provide sufficient and accurate diagnostic information to assist in resolving Android build failures, and how their effectiveness compares with the human-guided repair strategy developed in RQ2. Unlike previous studies that focused on static analysis or automation tools~\cite{li2017static}, our goal is not to enable LLMs to repair build failures automatically, but rather to evaluate whether the information and reasoning they generate can meaningfully support the repair process. To assess this capability, we randomly selected 15 representative applications from our dataset of 200 Android projects, categorized into three difficulty levels: minor issues, major issues, and unresolvable issues (five applications in each group, as defined in RQ2). For each case, the LLM (GPT-5) was provided with build logs, Gradle outputs, and error messages to generate explanations, identify root causes, and propose possible repair actions. Through this experiment, we aim to determine whether LLMs can replicate or surpass human performance in diagnosing and repairing build issues. We also examine their potential to address cases previously labeled as unresolvable, thereby evaluating the practical value of LLMs in augmenting Android build failure resolution.

{\textbf{LLM Prompt Design}}-The designed prompt~\cite{thisproject} defines a structured framework for guiding the LLM in Android build diagnosis and repair. It establishes an iterative, user-in-the-loop debugging workflow that emphasizes stepwise reasoning and transparency. The prompt requires the model to analyze provided build information—such as the GitHub repository, Gradle setup, and error logs—and respond in a standardized three-part format: (1) Suggested Action, specifying the proposed fix; (2) Command/Code/File Path, indicating where the change should be applied; and (3) Root Cause, explaining the rationale behind the issue. This structure ensures consistency, interpretability, and reproducibility in the model’s diagnostic responses.

{\textbf{Prompt Execution}}-The designed prompt operates through repeated interactions between the user and the LLM, reflecting the iterative nature of real-world Android debugging. The user provides the initial project context and build errors, after which the model analyzes the data and proposes one focused corrective action following the defined format. The user applies the fix, re-runs the build, and submits the updated error logs if new issues arise. This feedback loop continues until the project compiles successfully or no further solutions remain. Through this execution process, the LLM functions as a reasoning-based assistant, progressively refining its analysis and offering context-aware guidance across successive build failures.

{\textbf{Prompt Results}}-Using this prompt execution, we attempted to build 15 Android applications: five with minor issues, five with major issues, and five that were previously considered unresolvable. Among these, eight applications were successfully built with the assistance of the LLM (GPT-5)~\cite{GPT4o}, resulting in a success rate of 53.33\%. These results indicate that LLMs possess a notable capability to provide effective guidance in identifying and resolving build errors in Android projects. Specifically, all five applications with minor issues were successfully repaired following GPT-5’s suggestions, demonstrating its strength in addressing straightforward build problems. For the major issues, three out of five applications were resolved, suggesting that while the model can handle moderately complex scenarios, it still faces limitations when dealing with intricate dependency or configuration conflicts. None of the non-fixable cases were successfully compiled, indicating that the current state-of-the-art LLMs remain comparable to humans in these situations. If a human cannot resolve a build failure, the model is similarly unable to do so. During prompt execution, we observed that the LLM effectively examined Gradle and build logs to identify root causes such as outdated Gradle or Kotlin plugin versions, SDK mismatches, missing dependencies, and configuration inconsistencies, and then provided detailed, targeted remedies that conformed to the designed output format.

The two major build failures that were solvable by humans but not by the LLM were particularly revealing of the current limitations of such models. For example, the \emph{Clock app}~\cite{Clock} failed to compile due to a single outdated dependency that required replacing an entire library. The LLM repeatedly proposed incorrect fixes for the deprecated library instead of identifying a suitable modern alternative, ultimately entering a repetitive loop of ineffective suggestions. This case illustrates the model’s weakness in addressing issues that demand an understanding of the broader Android library ecosystem. Another example is the \emph{Water Reminder app}~\cite{water}, where the LLM correctly detected multiple outdated dependencies and recommended version upgrades. However, these upgrades introduced breaking API changes that caused secondary code-level compilation errors. The complete repair required refactoring the source code to comply with the updated APIs—a crucial follow-up step that the LLM neither anticipated nor performed. Together, these cases highlight the model’s difficulty in reasoning about the cascading effects of configuration-level fixes and its limited ability to handle problems that span both dependency management and source code adaptation.



\begin{tcolorbox}[size=title, opacityfill=0.15, enhanced, breakable]

\textbf{RQ3 Summary}: This study investigates the potential of LLMs to assist in diagnosing and resolving Android build failures. The findings indicate that LLMs can analyze build logs, identify underlying causes, and offer effective guidance for resolving common configuration and dependency issues, thereby improving the efficiency of the troubleshooting process. However, they still face challenges when dealing with complex or interdependent build failures that require deeper contextual understanding, such as recognizing deprecated libraries or performing code-level refactoring after dependency updates.

\end{tcolorbox}

\subsection{\textbf{RQ4: What project attributes influence the build of Android applications?}}

This research question examines the project attributes that influence whether an Android app can be built successfully, fails to build, or fails but can be fixed. Successfully building an Android app requires the proper coordination of multiple technical components, including the compatibility of JDK versions, Gradle configurations, and the system environment. Any mismatch among these components may lead to build errors, degraded performance, or complete build failure that studied in RQ1–RQ3. In addition to these build environment \& toolchain factors, this study investigates several project attributes—such as programming language, app size, last modified year, and user rating—that may not directly determine build success but can still exert an indirect influence on it. Understanding these project attributes helps explain why certain types of apps are more challenging to build than others and can inform prioritization strategies for selecting high-priority apps during software testing.

{\textbf{Programming languages}}- We investigated whether the programming language used in an Android project is a project attribute influencing build failures and fixability. 
In Android development, two primary programming languages are commonly used: \textit{Java} and \textit{Kotlin}. 
Java has been the original and officially supported language for Android development since the platform’s initial release in 2008. 
In contrast, Kotlin, developed by JetBrains, was officially supported by Google starting in 2017~\cite{Kotlin2017} and became the preferred language for Android development under the ``Kotlin-first'' policy announced in 2019~\cite{Kotlin2019}. 

\begin{figure}[h!]
  \centering
  \includegraphics[width=0.5\textwidth]{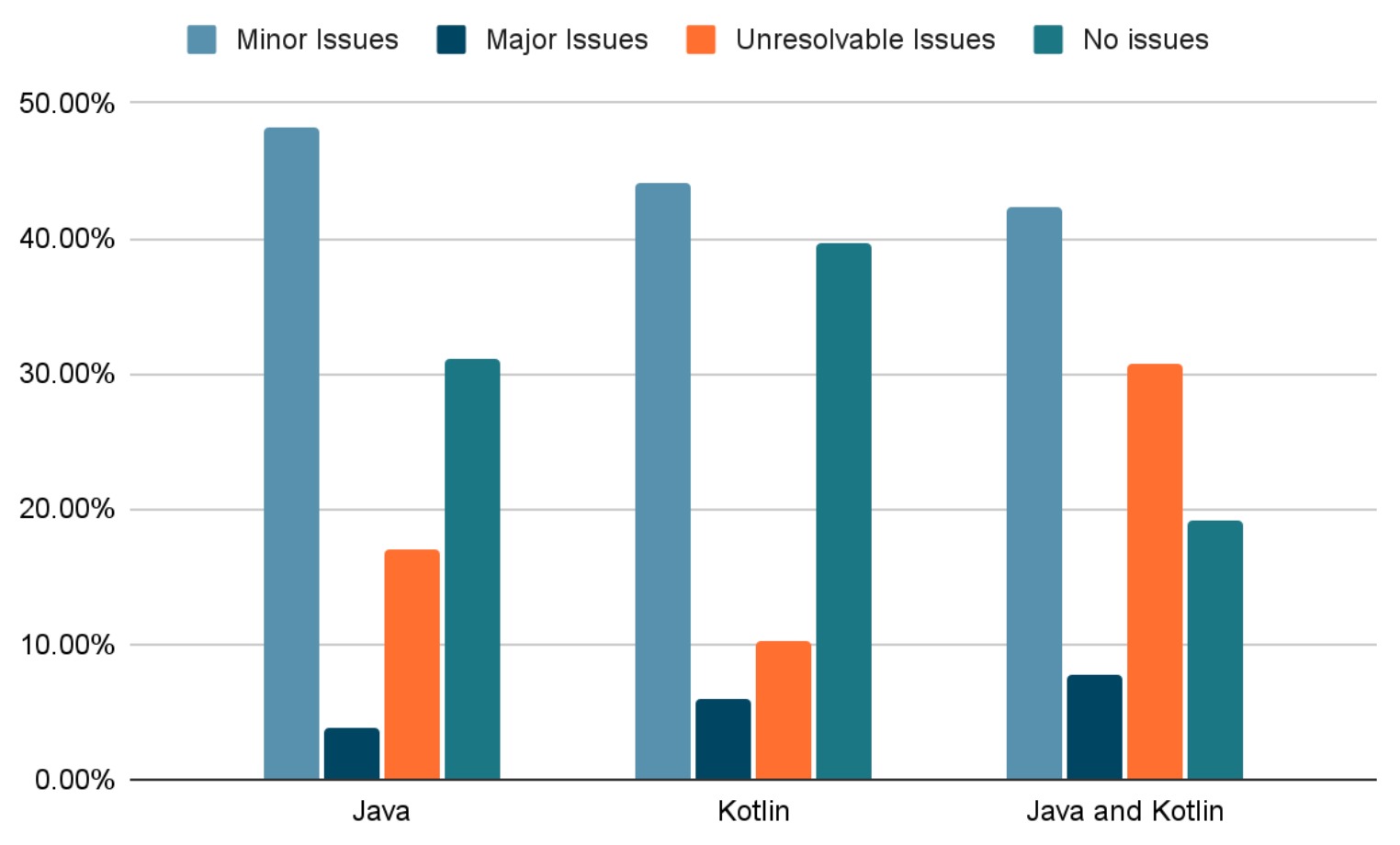}
  \caption{Number of issues faced in each language}
  \label{fig:language}
\end{figure}

In our dataset of 200 apps, we identified 106 written solely in Java, 68 written solely in Kotlin, and 26 written in a combination of both languages. 
We summarized the relationship between programming language and build results, as shown in Fig.~\ref{fig:language}. 
We found that 39.70\% of Kotlin apps had no build issues, which is significantly higher than Java apps (31.13\%). 
Moreover, 16.98\% of Java apps experienced build failures that could not be fixed, whereas only 10.29\% of Kotlin apps fell into this category. 
Although Kotlin apps exhibited a slightly higher rate of major issues (5.88\%) compared to Java apps (3.77\%), these issues were generally resolvable. Overall, our results suggest that Kotlin-based projects are more robust and easier to build successfully than those developed in Java and combination of both. 
This finding indicates that developers aiming for higher build success rates should consider adopting Kotlin for Android app development.

{\textbf{Last modified years}}- We investigated whether the last modified year of an Android project is a project attribute influencing build failures and fixability. Intuitively, one of the main reasons why an Android project fails to build is that some of its dependencies or libraries have become outdated or deprecated. When a project has not been updated for several years, it may rely on older SDK versions, deprecated APIs, or obsolete build tools that are no longer compatible with the current Android development environment. If this intuition holds, older apps are expected to be more difficult to compile and fix compared to newer ones. To verify this hypothesis, we conducted an empirical study to examine the relationship between the project’s last modified year and its build outcomes.

\begin{figure}[h!]
  \centering
  \includegraphics[width=0.5\textwidth]{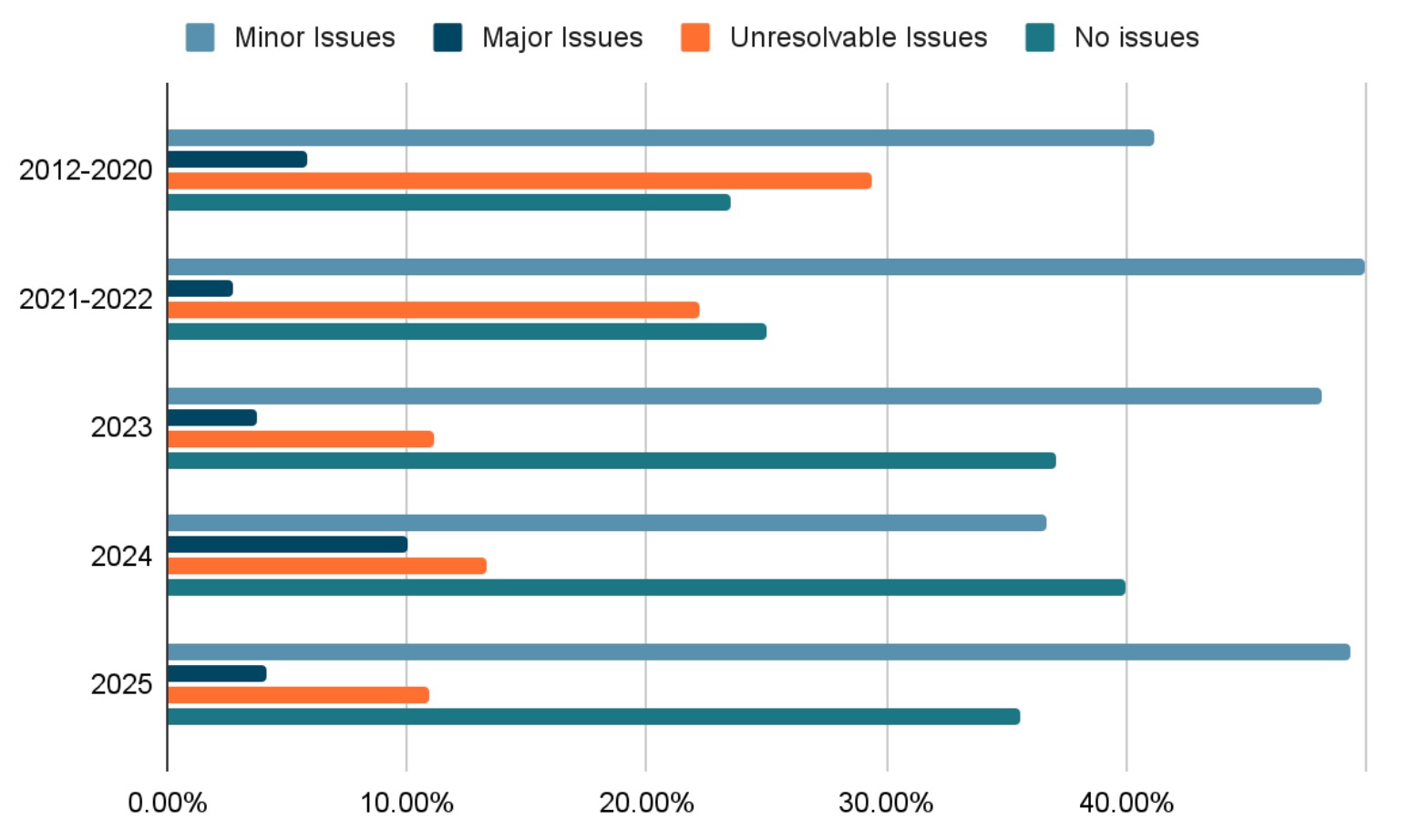}
  \caption{Year by year analysis of small,major,which cant be fixed and no issues occurence}
  \label{fig:year}
\end{figure}

We divided the apps into five groups by their last modified year to ensure a balanced sample distribution for statistical analysis: 34 apps from 2012–2020, 36 from 2021–2022, 27 from 2023, 30 from 2024, and 73 from 2025.
As shown in Fig.~\ref{fig:year}, the oldest group (2012–2020) had the lowest “no issues” build rate (23.52\%), followed by 2021–2022 (25\%). In contrast, newer apps (2023, 2024, and 2025) achieved substantially higher rates of 37.04\%, 40\%, and 35.62\%, respectively. Meanwhile, the oldest group also had the highest proportion of unresolvable issues (29.42\%), compared to only 10.96\% in 2025. These findings support our hypothesis that older apps, affected by outdated dependencies, are more likely to fail and harder to fix.

{\textbf{User rating}} — We further examined whether GitHub star counts influence build failures and fixability. Contrary to our initial expectation, the median ratings for apps in the 'Unresolvable issues,' 'No issues,' and 'Resolvable issues' categories were 3.1, 2.5, and 5.5, respectively. Interestingly, apps with unresolvable building issues had higher median ratings than issue-free apps. This may be because highly starred projects tend to be older and more complex, valued more for their functionality than their build maintainability.

\begin{figure}[h!]
  \centering
  \includegraphics[width=0.5\textwidth]{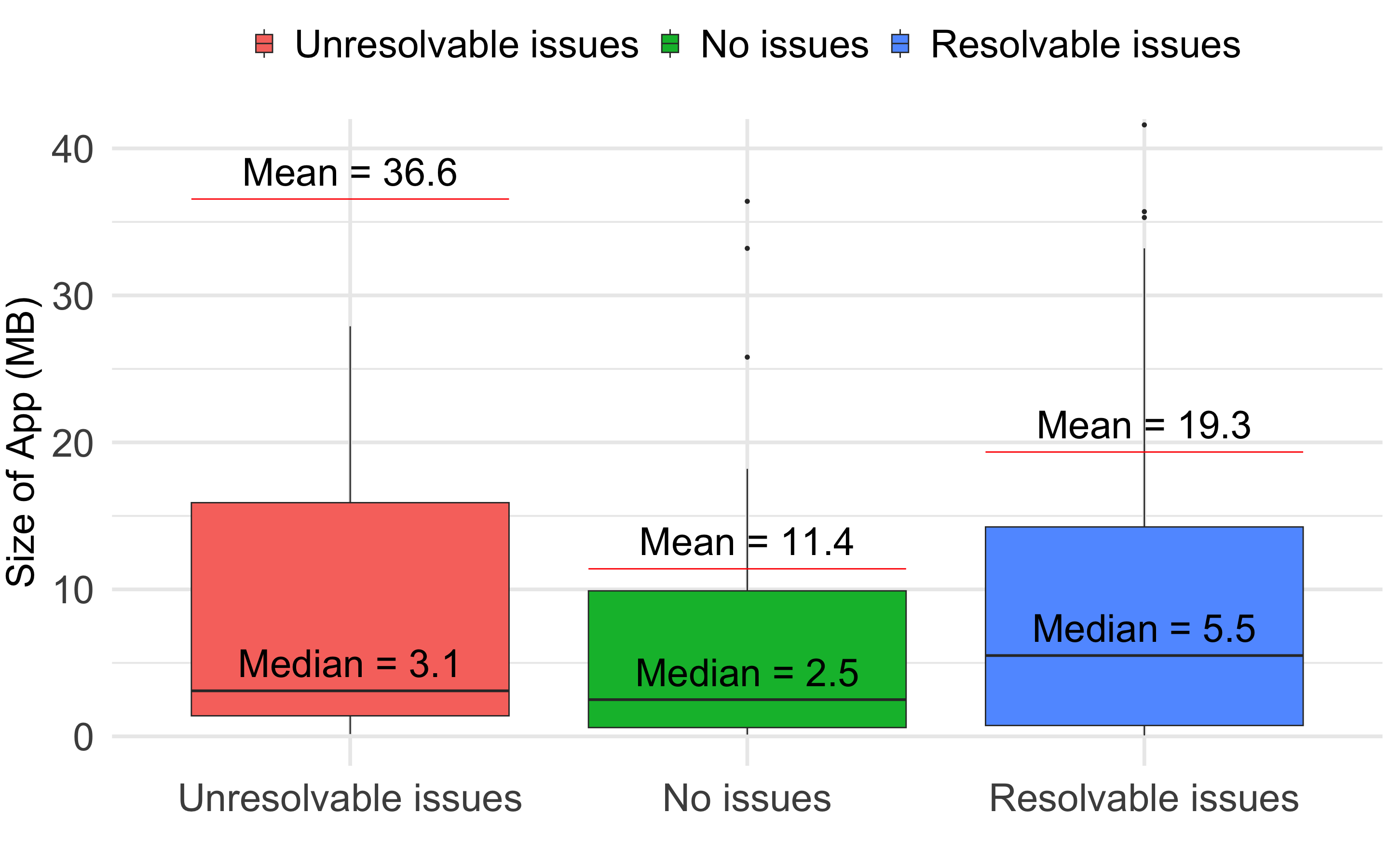}
  \caption{Boxplot for Issue Status vs Size of Apps}
  \label{fig:size}
\end{figure}

{\textbf{App size}} — Typically, larger applications are more complex and integrate more external dependencies. Consistent with the intuition that complexity increases build difficulty, our analysis in Fig.~\ref{fig:size} demonstrates that apps with unresolvable build issues had the largest average size (36.6 MB). For resolvable issues (which encompass both minor and major issues as defined in Section 3.2), apps showed a median size of 5.5 MB. Conversely, issue-free applications were the smallest, with a mean size of 11.4 MB and a median of 2.5 MB. These findings strongly indicate that larger applications generally pose more significant challenges for successful compilation.

\textbf{}

\vspace{-15pt}
\begin{tcolorbox}[size=title, opacityfill=0.15, enhanced, breakable]
\textbf{RQ4 Summary}:
Overall, our analysis shows that project attributes also influence Android app buildability. 
Kotlin-based and recently updated apps are more likely to build successfully, while older projects with outdated dependencies tend to fail or be harder to fix. 
App size exhibits a clear correlation—larger and more complex apps face more build issues. 
In contrast, user rating does not show a consistent relationship with build success, likely because highly-rated apps are often older and more complex. 
These findings suggest that project recency, complexity, and language choice indirectly affect the success of Android builds.

\end{tcolorbox}
\vspace{-15pt}


\section{Limitations and Validations}

This current study systematically analyzes compilation issues in 200 Android applications from GitHub and provides valuable insights, though it also has several limitations.
A key limitation is the inclusion of deprecated applications (e.g., \textit{NineOldAndroids}), which account for 33 cases (16.5\%) that could not be fixed. These negative examples were either abandoned or relied on outdated APIs (e.g., \textit{MyTodoList} suffers from Java heap issues), thereby increasing the proportion of unresolvable problems and potentially overstating the challenges faced in modern Android development.

Moreover, the exclusive use of open-source projects may limit generalizability to proprietary software, which often differs in maintenance strategies and resource allocation.
The dataset’s diversity in time span (2012–2025) and language distribution (Java 53\%, Kotlin 34\%, Java and Kotlin 13\%) also made build standardization more difficult. Legacy projects such as \textit{Face} recognition relied on outdated libraries and exhibited more dependency errors (42\%), whereas newer projects like \textit{Sunflower} had no issues. Although this temporal breadth enriches the analysis, it constrains the applicability of findings to modern Android environments (e.g., \textit{API 33+}). In addition, using Gradle as the sole build system excludes other compilation approaches and limits the scope of the study.

Despite the assistance of LLMs during error log analysis, manual examination of 200 applications could still introduce human error, particularly when classifying complex issues such as syntax/API errors (5\%). Furthermore, the lack of a control group makes it difficult to directly compare LLM-assisted methods with traditional approaches.

Nevertheless, the study demonstrates strong validity in several aspects. The large sample size and coverage of diverse domains—including file management (Amaze File Manager), education (Bookfinder), and media (Multimediaapp)—enhance external validity. The six-step strategy (cloning, syncing, error analysis, fixing, major resolution, recompilation) was thoroughly tested and achieved a {\textbf{75.56\%}} fix rate for resolvable cases, confirming its practical applicability and reproducibility.
The root cause classification is clearly defined (environment issues: 45\%, dependency and Gradle task errors: 42\%, configuration errors: 8\%, syntax/API errors: 5\%) and cross-verifiable with the Excel dataset, further strengthening internal validity.

Regarding construct validity, RQ3 demonstrates that LLMs can effectively interpret logs and propose solutions, as shown in Agecalculator and Florisboard. Theoretical validity is reinforced by the alignment of our findings with prior research (e.g., A Quantitative Study of Java Software Buildability ~\cite{sulir2016quantitative}) and by the high quality of metadata (language, time, issue types) in our dataset.
These layers of validity support the reliability of this study and show that, despite the inclusion of deprecated apps and reliance on open-source data, the proposed strategy and methodology remain practical and valuable for Android developers.

\section{Related Work}

Various research efforts have investigated the challenges of building software applications, particularly in the Java and Android ecosystems.
Among them, Hassan et al.~\cite{hassan2017automatic} conducted one of the large-scale empirical studies on the feasibility of automatic Java project building from open-source repositories.
They systematically examined 200 popular Java projects on GitHub using Ant, Maven, and Gradle, and found that nearly half of the projects failed to build using default commands—primarily due to environment mismatches, non-default build parameters, and missing dependencies.
While their study provides valuable insights into general Java build automation, it does not address Android-specific challenges, such as Android Gradle Plugin (AGP) compatibility, SDK version fragmentation, and mobile-oriented dependency management. Sulír and Porubän ~\cite{sulir2016quantitative} conducted a large-scale quantitative study on the buildability of Java projects, revealing that approximately 38\% of open-source Java systems failed to build, with dependency and compilation errors being the most common causes. While their study provided a foundational understanding of build failures in the general Java ecosystem, it excluded Android projects and did not explore practical repair workflows. For instance, issues such as hard-coded paths are rarely relevant to Android projects, whereas Android-specific problems like manifest misconfigurations and Android Gradle Plugin (AGP) version conflicts are often overlooked.

Some of tasks are specifically focusing on Android compiling issues, which are more related to our research. Zhang et al. ~\cite{zhang2016android} proposed an automated approach to extract and visualize build dependency structures in large-scale systems such as Android OS. Their method monitors the GNU Make process to reconstruct command invocation trees and inter-artifact dependencies, providing insights into build efficiency and dependency evolution between Android 5 and 6. Jha et al., ~\cite{jha2017developer} conducted one of the large-scale empirical analyses on manifest files mistakes in Android applications. By analyzing apps’ manifest files using a rule-based static analysis tool (ManifestInspector), they identified misconfigurations across distinct error types, highlighting common developer mistakes in security permissions, component declarations, and UI configurations. Recent work by Liu et al.~\cite{liu2024understanding} conducted a large-scale empirical study on the evolution and quality of Android build systems. Their study highlights the dominance of Gradle and reveals that only about one-third of projects can be automatically built due to configuration inconsistencies and deprecated build scripts. 

These works primarily emphasize structural or statistical analyses rather than practical resolution. In contrast, our study represents the most recent and comprehensive effort to empirically diagnose and repair build failures across 200 Android applications over an extended time span from 2012 to 2025. We not only categorize root causes and introduce a systematic repair strategy augmented by LLMs for automated diagnosis, but also provide detailed solution links, logs, and illustrative screenshots for each case. This ensures that our dataset and methodology can serve as a reproducible and extensible resource for future researchers investigating Android build reliability and automation. Additionally, this study performs a temporal and linguistic analysis of Android projects’ buildability. By characterizing these factors, our work offers the first ecosystem-level perspective on the temporal and linguistic dynamics of Android compilation.

Research on Android applications and software engineering is increasingly extending beyond application building to leverage LLMs for automation and performance enhancement.
For instance, AdbGPT~\cite{feng2024prompting} and BugRepro~\cite{yin2025bugrepro} employ LLMs to reproduce Android bugs from natural-language descriptions. PG-TD~\cite{zhang2023planning} applies LLMs to generate and refine program code, while GPTDroid~\cite{liu2024make} focuses on synthesizing Android test sequences. LLMDroid~\cite{wang2025llmdroid} further leverages LLMs to increase the code coverage of Android applications.
Unlike these studies, which primarily target code generation, testing, or bug reproduction, this work tackles a fundamental bottleneck in the Android ecosystem, the reliability and feasibility of application compilation.

Some Android testing approaches~\cite{ju2024study, zhao2024dinodroid, vuong2019semantic, mao2016sapienz, su2017guided} and bug report reproduction techniques~\cite{zhang2023automatically, zhao2019recdroid, zhao2022recdroid+, wang2024feedback, feng2024prompting} rely on dynamic execution and runtime code coverage collection, typically using tools such as Emma~\cite{Emma} and JaCoCo~\cite{Jacoco}.
All these testing and reproduction tasks require the compiled Android applications as their target software.
Therefore, constructing a compilable and executable dataset of Android apps is a critical prerequisite for these studies.

\vspace{-12 pt}

\section{Conclusion}

This study investigates the technical bottlenecks in building Android applications. With the rapid growth of the Android ecosystem, building has become a complex process involving dependency management, environment consistency, toolchain compatibility, and language-specific factors. We analyzed 200 open-source Android projects (Java and Kotlin) and identified four main causes of build failures: environment issues, dependency and Gradle task errors, configuration problems, and syntax/API incompatibilities. Among these, 65 projects built successfully without modification, 92 required minor fixes, 10 required major fixes, and 33 were unresolvable. A repair strategy enabled 102 of 135 failing projects to be fixed (75.56\%). GPT-5 shows strong potential for software build and maintenance. Java projects were more prone to environment and dependency issues, while Kotlin projects showed better stability. 

In future research, we plan to leverage our curated human-labeled dataset and the continuously generated build data to fine-tune a local LLM specialized for Android build diagnosis and repair.

\bibliographystyle{IEEEtran}
\bibliography{mybiblography}

@misc{Calculator,
title = {{Calculator}},
year={2016},
howpublished={\url{https://github.com/anubhavshrimal/CalculatorApp/tree/b36c1ccc4602e8c260253c35cecf1921df73baa6}},
}

@misc{Clock,
title = {{Clock App}},
year={2024},
howpublished={\url{https://github.com/SimpleMobileTools/Simple-Clock/tree/28b9866ff7baafc8447ea3075baa7720f45926ec}},
}

@misc{water,
title = {{Water Reminder}},
year={2021},
howpublished={\url{https://github.com/KeyurDiwan/Water-Reminder/tree/3f35db41555ea16f3259ca20fab5d8ed977fa401}},
}

@misc{Cryptography,
title = {{Cryptography}},
year={2022},
howpublished={\url{https://github.com/Shijas-T/Cryptography-App/tree/bd11ed2bc2abe89eeeefa364a5a8d1aa89f37523}},
}

@misc{Meme,
title = {{MemeShare}},
year={2020},
howpublished={\url{https://github.com/Anuj-Kumar-Sharma/Meme-Share/tree/e799ce0f68129dd97b7408b03cc2142a03145683 }},
}

@misc{Voice,
title = {{VoiceRecord}},
year={2024},
howpublished={\url{https://github.com/SimpleMobileTools/Simple-Voice-Recorder/tree/db0d3e30b7088b9bff5b385c32a22b432673c351}},
}

@misc{Kotlin2017,
title = {{Kotlin2017}},
year={2025},
howpublished={\url{https://android-developers.googleblog.com/2017/05/android-announces-support-for-kotlin.html}},
}

@misc{Kotlin2019,
title = {{Kotlin2019}},
year={2025},
howpublished={\url{https://developer.android.com/kotlin/first}},
}

@misc{using,
title = {{Using Basic Room Database Master}},
year={2023},
howpublished={\url{https://github.com/mehdisahraeei/using-basic-room-database/tree/e532111c1dcac3e8e378b266c3d293d4d12c0847}},
}

@misc{Android,
title = {{Android}},
year={2025},
howpublished={\url{https://github.com/MihaiCristianCondrea/Android-Studio-Tutorials-Java-Edition/tree/6906dd03b3ed067d5a0beb82e8bd119c86814164}},
}

@misc{Anime,
title = {{LikeAnimation}},
year={2018},
howpublished={\url{https://github.com/frogermcs/LikeAnimation/tree/0aa95cc6ad6a75438e76e14244fd9e266d5906ef }},
}

@misc{Age,
title = {{Age for minutes kt}},
year={2022},
howpublished={\url{https://github.com/EleoXDA/Age_For_Minutes_KT/tree/da8db77610b3e344b9b13b5aeb840b41ad9807a0}},
}

@book{craig2015learn,
  title={Learn Android Studio: build Android apps quickly and effectively},
  author={Craig, Clifton and Gerber, Adam},
  year={2015},
  publisher={Apress}
}

@article{li2017static,
  title={Static analysis of android apps: A systematic literature review},
  author={Li, Li and Bissyand{\'e}, Tegawend{\'e} F and Papadakis, Mike and Rasthofer, Siegfried and Bartel, Alexandre and Octeau, Damien and Klein, Jacques and Traon, Le},
  journal={Information and Software Technology},
  volume={88},
  pages={67--95},
  year={2017},
  publisher={Elsevier}
}

@misc{thisproject,
title = {{Github project}},
year={2025},
howpublished={\url{https://github.com/lunarinus4/Android-build}},
}

@misc{github,
title = {{Github}},
year={2025},
howpublished={\url{https://github.com}},
}

@misc{bitbucket,
title = {{Bitbucket}},
year={2025},
howpublished={\url{https://bitbucket.org/product/}},
}

@article{sulir2020large,
  title={Large-scale dataset of local java software build results},
  author={Sul{\'\i}r, Mat{\'u}{\v{s}} and Ba{\v{c}}{\'\i}kov{\'a}, Michaela and Madeja, Matej and Chodarev, Sergej and Juh{\'a}r, J{\'a}n},
  journal={Data},
  volume={5},
  number={3},
  pages={86},
  year={2020},
  publisher={MDPI}
}

@inproceedings{zhang2016android,
  title={Android build dependency analysis},
  author={Zhang, Bo and Tenev, Vasil and Becker, Martin},
  booktitle={2016 IEEE 24th International Conference on Program Comprehension (ICPC)},
  pages={1--4},
  year={2016},
  organization={IEEE}
}

@misc{googleplaylink,
title = {{Google Play Data}},
year={2020},
howpublished={\url{https://www.statista.com/statistics/266210/number-of-available-applications-in-the-google-play-store}},
}

@article{zhang2023planning,
  title={Planning with large language models for code generation},
  author={Zhang, Shun and Chen, Zhenfang and Shen, Yikang and Ding, Mingyu and Tenenbaum, Joshua B and Gan, Chuang},
  journal={arXiv preprint arXiv:2303.05510},
  year={2023}
}

@article{yin2025bugrepro,
  title={BugRepro: Enhancing Android Bug Reproduction with Domain-Specific Knowledge Integration},
  author={Yin, Hongrong and Huang, Jinhong and Li, Yao and Dong, Yunwei and Zhang, Tao},
  journal={arXiv preprint arXiv:2505.14528},
  year={2025}
}

@article{wang2025llmdroid,
  title={LLMDroid: Enhancing Automated Mobile App GUI Testing Coverage with Large Language Model Guidance},
  author={Wang, Chenxu and Liu, Tianming and Zhao, Yanjie and Yang, Minghui and Wang, Haoyu},
  journal={Proceedings of the ACM on Software Engineering},
  volume={2},
  number={FSE},
  pages={1001--1022},
  year={2025},
  publisher={ACM New York, NY, USA}
}

@inproceedings{liu2024make,
  title={Make llm a testing expert: Bringing human-like interaction to mobile gui testing via functionality-aware decisions},
  author={Liu, Zhe and Chen, Chunyang and Wang, Junjie and Chen, Mengzhuo and Wu, Boyu and Che, Xing and Wang, Dandan and Wang, Qing},
  booktitle={Proceedings of the IEEE/ACM 46th International Conference on Software Engineering},
  pages={1--13},
  year={2024}
}

@article{liu2024understanding,
  title={Understanding the quality and evolution of Android app build systems},
  author={Liu, Pei and Li, Li and Liu, Kui and McIntosh, Shane and Grundy, John},
  journal={Journal of Software: Evolution and Process},
  volume={36},
  number={5},
  pages={e2602},
  year={2024},
  publisher={Wiley Online Library}
}

@inproceedings{jha2017developer,
  title={Developer mistakes in writing android manifests: An empirical study of configuration errors},
  author={Jha, Ajay Kumar and Lee, Sunghee and Lee, Woo Jin},
  booktitle={2017 IEEE/ACM 14th International Conference on Mining Software Repositories (MSR)},
  pages={25--36},
  year={2017},
  organization={IEEE}
}

@inproceedings{sulir2016quantitative,
  title={A quantitative study of java software buildability},
  author={Sul{\'\i}r, Mat{\'u}{\v{s}} and Porub{\"a}n, Jaroslav},
  booktitle={Proceedings of the 7th International Workshop on Evaluation and Usability of Programming Languages and Tools},
  pages={17--25},
  year={2016}
}

@inproceedings{hassan2017automatic,
  title={Automatic building of java projects in software repositories: A study on feasibility and challenges},
  author={Hassan, Foyzul and Mostafa, Shaikh and Lam, Edmund SL and Wang, Xiaoyin},
  booktitle={2017 ACM/IEEE International Symposium on Empirical Software Engineering and Measurement (ESEM)},
  pages={38--47},
  year={2017},
  organization={IEEE}
}

@misc{emma,
title = {{Emma}},
year={2025},
howpublished={\url{m http://emma.sourceforge.net/}},
}

@misc{jacoco,
title = {{Jacoco}},
year={2025},
howpublished={\url{m https://github.com/arturdm/jacoco-android-gradle-plugin}},
}

@misc{GPT4o,
year={2024},
title = {GPT-4o},
howpublished={\url{https://openai.com/index/hello-gpt-4o/}},
}

@inproceedings{wen2024autodroid,
  title={Autodroid: Llm-powered task automation in android},
  author={Wen, Hao and Li, Yuanchun and Liu, Guohong and Zhao, Shanhui and Yu, Tao and Li, Toby Jia-Jun and Jiang, Shiqi and Liu, Yunhao and Zhang, Yaqin and Liu, Yunxin},
  booktitle={Proceedings of the 30th Annual International Conference on Mobile Computing and Networking},
  pages={543--557},
  year={2024}
}

@inproceedings{wang2023empirical,
  title={An Empirical Study of Regression Testing for Android Apps in Continuous Integration Environment},
  author={Wang, Dingbang and Zhao, Yu and Xiao, Lu and Yu, Tingting},
  booktitle={2023 ACM/IEEE International Symposium on Empirical Software Engineering and Measurement (ESEM)},
  pages={1--11},
  year={2023},
  organization={IEEE}
}

@article{zhao2024dinodroid,
  title={Dinodroid: Testing android apps using deep q-networks},
  author={Zhao, Yu and Harrison, Brent and Yu, Tingting},
  journal={ACM Transactions on Software Engineering and Methodology},
  volume={33},
  number={5},
  pages={1--24},
  year={2024},
  publisher={ACM New York, NY}
}

@inproceedings{su2017guided,
  title={Guided, stochastic model-based GUI testing of Android apps},
  author={Su, Ting and Meng, Guozhu and Chen, Yuting and Wu, Ke and Yang, Weiming and Yao, Yao and Pu, Geguang and Liu, Yang and Su, Zhendong},
  booktitle={Proceedings of the 2017 11th joint meeting on foundations of software engineering},
  pages={245--256},
  year={2017}
}

@inproceedings{mao2016sapienz,
  title={Sapienz: Multi-objective automated testing for android applications},
  author={Mao, Ke and Harman, Mark and Jia, Yue},
  booktitle={Proceedings of the 25th international symposium on software testing and analysis},
  pages={94--105},
  year={2016}
}

@inproceedings{vuong2019semantic,
  title={Semantic Analysis for Deep Q-Network in Android GUI Testing.},
  author={Vuong, Thi Anh Tuyet and Takada, Shingo},
  booktitle={SEKE},
  pages={123--170},
  year={2019}
}

@inproceedings{feng2024prompting,
  title={Prompting is all you need: Automated android bug replay with large language models},
  author={Feng, Sidong and Chen, Chunyang},
  booktitle={Proceedings of the 46th IEEE/ACM International Conference on Software Engineering},
  pages={1--13},
  year={2024}
}

@inproceedings{ju2024study,
  title={A study of using multimodal llms for non-crash functional bug detection in android apps},
  author={Ju, Bangyan and Yang, Jin and Yu, Tingting and Abdullayev, Tamerlan and Wu, Yuanyuan and Wang, Dingbang and Zhao, Yu},
  booktitle={2024 31st Asia-Pacific Software Engineering Conference (APSEC)},
  pages={61--70},
  year={2024},
  organization={IEEE}
}

@inproceedings{wang2024feedback,
  title={Feedback-driven automated whole bug report reproduction for android apps},
  author={Wang, Dingbang and Zhao, Yu and Feng, Sidong and Zhang, Zhaoxu and Halfond, William GJ and Chen, Chunyang and Sun, Xiaoxia and Shi, Jiangfan and Yu, Tingting},
  booktitle={Proceedings of the 33rd ACM SIGSOFT International Symposium on Software Testing and Analysis},
  pages={1048--1060},
  year={2024}
}

@article{zhao2022recdroid+,
  title={Recdroid+: Automated end-to-end crash reproduction from bug reports for android apps},
  author={Zhao, Yu and Su, Ting and Liu, Yang and Zheng, Wei and Wu, Xiaoxue and Kavuluru, Ramakanth and Halfond, William GJ and Yu, Tingting},
  journal={ACM Transactions on Software Engineering and Methodology (TOSEM)},
  volume={31},
  number={3},
  pages={1--33},
  year={2022},
  publisher={ACM New York, NY}
}

@inproceedings{zhao2019recdroid,
  title={Recdroid: automatically reproducing android application crashes from bug reports},
  author={Zhao, Yu and Yu, Tingting and Su, Ting and Liu, Yang and Zheng, Wei and Zhang, Jingzhi and Halfond, William GJ},
  booktitle={2019 IEEE/ACM 41st International Conference on Software Engineering (ICSE)},
  pages={128--139},
  year={2019},
  organization={IEEE}
}

@inproceedings{zhang2023automatically,
  title={Automatically reproducing android bug reports using natural language processing and reinforcement learning},
  author={Zhang, Zhaoxu and Winn, Robert and Zhao, Yu and Yu, Tingting and Halfond, William GJ},
  booktitle={Proceedings of the 32nd ACM SIGSOFT International Symposium on Software Testing and Analysis},
  pages={411--422},
  year={2023}
}

@article{madeja2021automating,
  title={Automating test case identification in java open source projects on github},
  author={Madeja, Matej and Porub{\"a}n, Jaroslav and Ba{\v{c}}{\'\i}kov{\'a}, Michaela and Sul{\'\i}r, Mat{\'u}{\v{s}} and Juh{\'a}r, J{\'a}n and Chodarev, Sergej and Gurb{\'a}l', Filip},
  journal={arXiv preprint arXiv:2102.11678},
  year={2021}
}


\end{document}